\documentclass[prd,amsmath,amssymb,superscriptaddress,preprintnumbers,twocolumn,nofootinbib,10pt]{revtex4-1}
\usepackage{graphicx}
\usepackage{dcolumn}
\usepackage{bm}
\usepackage{amssymb}
\usepackage{latexsym}
\usepackage{booktabs}
\usepackage{amsmath}
\usepackage{multirow}
\usepackage{url}
\usepackage{footnote}
\usepackage{float}
\usepackage{threeparttable}
\usepackage[colorlinks=true, linkcolor=blue, citecolor=blue]{hyperref}
\usepackage[bottom]{footmisc}

\usepackage[normalem]{ulem}
\usepackage{color}
\usepackage{array}
\usepackage{enumerate}
\usepackage{adjustbox}

\usepackage{makecell}
\usepackage{diagbox}
\usepackage{epstopdf}
\usepackage{epsfig}
\usepackage{longtable}
\usepackage{supertabular}
\usepackage{algorithm}
\usepackage{pifont}
\usepackage{algorithmic}
\usepackage{changepage}
\usepackage{setspace}
\begin{document}

\title{Testing the cosmic distance duality relation with baryon acoustic oscillations and supernovae data} 
\author{Tian-Nuo Li}
\affiliation{Key Laboratory of Cosmology and Astrophysics (Liaoning Province), College of Sciences, Northeastern University, Shenyang 110819, China}

\author{Guo-Hong Du}
\affiliation{Key Laboratory of Cosmology and Astrophysics (Liaoning Province), College of Sciences, Northeastern University, Shenyang 110819, China}

\author{Peng-Ju Wu}
\affiliation{School of Physics, Ningxia University, Yinchuan 750021, China}

\author{Jing-Zhao Qi}
\affiliation{Key Laboratory of Cosmology and Astrophysics (Liaoning Province), College of Sciences, Northeastern University, Shenyang 110819, China}

\author{Jing-Fei Zhang}\thanks{Corresponding author}\email{ jfzhang@mail.neu.edu.cn}
\affiliation{Key Laboratory of Cosmology and Astrophysics (Liaoning Province), College of Sciences, Northeastern University, Shenyang 110819, China}

\author{Xin Zhang}\thanks{Corresponding author}\email{zhangxin@mail.neu.edu.cn}
\affiliation{Key Laboratory of Cosmology and Astrophysics (Liaoning Province), College of Sciences, Northeastern University, Shenyang 110819, China}
\affiliation{MOE Key Laboratory of Data Analytics and Optimization for Smart Industry, Northeastern University, Shenyang 110819, China}
\affiliation{National Frontiers Science Center for Industrial Intelligence and Systems Optimization, Northeastern University, Shenyang 110819, China}

\begin{abstract}

{One of the most fundamental relationships in modern cosmology is the cosmic distance duality relation (CDDR), which describes the relationship between the angular diameter distance ($D_{\rm A}$) and the luminosity distance ($D_{\rm L}$), and is expressed as $\eta(z)=D_{\rm L}(z)(1+z)^{-2}/D_{\rm A}(z)=1$. In this work, we conduct a comprehensive test of the CDDR by combining baryon acoustic oscillation (BAO) data from the SDSS and DESI surveys with type Ia supernova (SN) data from PantheonPlus and DESY5. We utilize an artificial neural network approach to match the SN and BAO data at the same redshift. To explore potential violations of the CDDR, we consider three different parameterizations: (i) $\eta(z)=1+\eta_0z$; (ii) $\eta(z)=1+\eta_0z/(1+z)$; (iii) $\eta(z)=1+\eta_0\ln(1+z)$. 
Our results indicate that the calibration of the SN absolute magnitude $M_{\rm B}$ plays a crucial role in testing potential deviations from the CDDR, as there exists a significant negative correlation between $\eta_0$ and $M_{\rm B}$. For PantheonPlus analysis, when $M_{\rm B}$ is treated as a free parameter, no evidence of CDDR violation is found. In contrast, fixing $M_{\rm B}$ to the $M_{\rm B}^{\rm D20}$ prior with $-19.230\pm0.040$ mag leads to a deviation at approximately the $2\sigma$ level, while fixing $M_{\rm B}$ to the $M_{\rm B}^{\rm B23}$ prior with $-19.396\pm0.016$ mag remains in agreement with the CDDR. Furthermore, overall analyses based on the SDSS+DESY5 and DESI+DESY5 data consistently show no evidence of deviation from the CDDR.}

\end{abstract}
\maketitle

\section{Introduction}
The cosmic distance duality relation (CDDR), also known as the Etherington relation \cite{Etherington1933}, is a fundamental relationship in modern cosmology. This relationship is predicated on three fundamental assumptions: the description of spacetime by the metric theory of gravity, the propagation of photons along null geodesics, and the conservation of the photon number \cite{Bassett:2003vu}. Specifically, for a given redshift, the relation can be expressed as $D_{\rm L}(z) = (1+z)^2 D_{\rm A}(z)$, which establishes a connection between two critical measures of cosmic distance for an astronomical object: the luminosity distance ($D_{\rm L}$) and the angular diameter distance ($D_{\rm A}$). Therefore, testing the validity of the CDDR is essential for verifying the theories of gravity and the foundations of modern cosmology \cite{Avgoustidis2010,Holanda2011}.

On the one hand, violations of the CDDR can arise in certain modified theories of gravity in which spacetime is no longer described by a pseudo-Riemannian manifold \cite{Santana:2017zvy,Azevedo:2021npm}. For example, in torsion-based theories, the connection is not purely Levi-Civita but includes torsional terms \cite{Cai:2015emx}; similarly, in energy-momentum squared gravity, the field equations include higher-order terms of the energy-momentum components \cite{Cipriano:2024jng}. Moreover, violations of the CDDR may also arise from photons not traveling along null geodesics or from the non-conservation of photon number \cite{Uzan:2004my}. For example, the non-conservation of the photon number can be associated with the presence of certain opacity sources and non-standard mechanisms, such as axion-photon conversion induced by intergalactic magnetic fields \cite{Bassett:2003vu,Avgoustidis:2009ai}, or scalar fields with a non-minimal multiplicative coupling to the electromagnetic Lagrangian \cite{Holanda:2016ala}. On the other hand, violations of the CDDR may indicate the presence of systematic errors in distance measurements. It is well known that there exists a significant discrepancy in the current measurements of the Hubble constant ($H_0$): the SH0ES collaboration, using the local distance ladder method based on Cepheid-calibrated supernovae, reports a value of $H_0=73.04\pm1.04$ ${\rm km}~{\rm s}^{-1}~{\rm Mpc}^{-1}$ \cite{Riess:2021jrx}, whereas the Planck collaboration, assuming a $\Lambda$ cold dark matter ($\Lambda$CDM) cosmology and based on the cosmic microwave background (CMB) data, infers a value of $H_0=67.36\pm0.54$ ${\rm km}~{\rm s}^{-1}~{\rm Mpc}^{-1}$ \cite{Planck:2018vyg}. The discrepancy has reached statistical significance of more than 5$\sigma$, which is referred to as the Hubble tension problem.\footnote{In recent years, the Hubble tension has been extensively discussed \cite{Bernal:2016gxb,DiValentino:2017iww,Guo:2018ans,Vagnozzi:2019ezj,DiValentino:2019ffd,Cai:2021wgv,Escudero:2022rbq,Zhao:2022yiv,James:2022dcx,Vagnozzi:2023nrq,Zhang:2023gye,Pierra:2023deu,Jin:2023sfc,Huang:2024erq,Xiao:2024nmi,Han:2025fii,Song:2025ddm,Zhang:2024rra,Dong:2024bvw,Jin:2025dvf}; see also Refs.~\cite{DiValentino:2021izs,Kamionkowski:2022pkx} for reviews.} This significant discrepancy may arise from systematic effects in observations; it may also point towards new physics beyond the $\Lambda$CDM model. Therefore, testing the validity of the CDDR serves as a powerful tool for probing potential deviations from the $\Lambda$CDM paradigm.

In recent years, a considerable amount of research has employed various astronomical observations to test the CDDR \cite{Uzan:2004my,DeBernardis:2006ii,Li:2011exa,Nair:2011dp,Holanda:2010vb,Liang:2011gm,Meng:2011nt,Khedekar:2011gf,Goncalves:2011ha,Lima:2011ye,Cardone:2012vd,Yang:2013coa,Wu:2015ixa,Ma:2016bjt,Holanda:2016msr,More:2016fca,Rana:2017sfr,Li:2017zrx,Lin:2018qal,Qi:2019spg,Holanda:2019vmh,Zhou:2020moc,Qin:2021jqy,Bora:2021cjl,Mukherjee:2021kcu,Liu:2021fka,Xu:2022zlm,Tonghua:2023hdz,Yang:2024icv,Tang:2024zkc,Jesus:2024nrl,Qi:2024acx,Alfano:2025gie}. Type Ia supernova (SN) have been widely used to determine the luminosity distance, whereas the angular diameter distance, is typically obtained through multiple observational techniques, including (but not limited to) the Sunyaev--Zeldovich effect and gas mass fraction measurements in galaxy clusters \cite{Holanda:2010vb,Li:2011exa}, strong gravitational lensing systems \cite{Liao:2022gde,Ruan:2018dls,Qi:2024acx}, the angular size of ultra-compact radio sources \cite{Li:2017zrx}, and baryon acoustic oscillations (BAO) \cite{Wu:2015ixa}. In particular, SN and BAO observations play an important role in testing the CDDR \cite{Wu:2015ixa,Xu:2020fxj,Wang:2024rxm,Xu:2022zlm}. For example, Ref.~\cite{Wu:2015ixa} tested the CDDR by comparing the SN data from the SCP Union2.1 compilation with five BAO measurements, and concluded that the high precision of BAO measurements makes them an effective tool for validating the CDDR. Reference~\cite{Wang:2024rxm} tested the CDDR relation by utilizing SN data from the Pantheon sample and transverse BAO measurements from the Sloan Digital Sky Survey (SDSS), demonstrating that transverse BAO measurements can serve as a powerful tool for validating the CDDR relation.

Recently, the Dark Energy Spectroscopic Instrument (DESI) publicly released its second data release (DR2), which includes measurements of BAO from over 14 million extragalactic sources, spanning the redshift range $0.295 \leq z \leq 2.330$. This release has marked the beginning of a new era in high-precision BAO measurements. The DESI DR2 BAO data, combined with SN data from PantheonPlus, Union3, and DESY5, along with CMB data from Planck and ACT, provide more stringent constraints on the dark energy equation of state, revealing a $2.8\sigma-4.2\sigma$ preference for dynamical dark energy \cite{DESI:2025zgx}. DESI BAO data have attracted a significant number of researchers seeking to constrain various aspects of cosmological physics, such as dark energy \cite{Li:2024qso,Giare:2024gpk,Dinda:2024ktd,Colgain:2024xqj,Escamilla:2024ahl,Colgain:2024ksa,Sabogal:2024yha,Colgain:2024mtg,Li:2024qus,Li:2024bwr,Wang:2024dka,Giare:2024smz,Huang:2025som,Li:2025owk,Wu:2025wyk,Li:2025ula,Li:2025ops,Barua:2025ypw,Yashiki:2025loj,Ling:2025lmw,Goswami:2025uih,Yang:2025boq,Colgain:2025nzf,Pang:2025lvh,You:2025uon,Ozulker:2025ehg,Cheng:2025lod,Pan:2025qwy,Qiang:2025cxp,Liu:2025myr,Gialamas:2025pwv,Colgain:2025fct,Wu:2025vfs}, non-cold dark matter \cite{Yang:2025ume,Chen:2025wwn,Wang:2025zri,Kumar:2025etf,Abedin:2025dis, Khoury:2025txd,Araya:2025rqz,Li:2025eqh,Li:2025dwz,Paliathanasis:2025xxm}, neutrinos \cite{Du:2024pai,Jiang:2024viw,Du:2025iow,Feng:2025mlo,RoyChoudhury:2025dhe,Du:2025xes,Zhou:2025nkb}, and modified gravity \cite{Cai:2025mas,Li:2025cxn,Odintsov:2025jfq,Nojiri:2025low}. Additionally, it is essential to revisit the CDDR in the context of the most recent high-precision BAO data from DESI, as emphasized by several studies based on the first data release of DESI \cite{Favale:2024sdq,Yang:2025qdg,Keil:2025ysb,Teixeira:2025czm} and more recent analyses using DESI DR2 data \cite{Zhang:2025qbs,Wang:2025gus,Santos:2025gjf, Afroz:2025iwo}. For example, Ref.~\cite{Favale:2024sdq} investigated the consistency of the CDDR by combining SDSS and DESI mixed BAO data with SN data, reporting indications of a potential deviation at approximately the $2\sigma$ level. The authors attributed this deviation to possible tensions between the SN and BAO data, or potentially internal inconsistencies within the BAO data subsets.

However, a major challenge in testing the CDDR using the latest DESI DR2 data is the difficulty of obtaining matched angular diameter distance and luminosity distance at identical redshifts, as most observational datasets provide these quantities at different redshift points. To address this issue, several methods for reconstructing the distance have been proposed, such as linear and polynomial fitting~\cite{Holanda:2015zpz} and Gaussian process~\cite{Hees:2014lfa}. Specifically, Ref.~\cite{Wang:2019vxv} proposed a non-parametric approach based on an artificial neural network (ANN) to reconstruct functions directly from observational data. In recent years, several studies have applied the ANN method to test the CDDR; see, e.g., Refs. \cite{Liu:2021fka,Xu:2022zlm,Tang:2022ykd,Qi:2024acx,Yang:2025qdg}.

In this work, our primary motivation is to test the validity of the CDDR by utilizing the latest DESI DR2 BAO and DESY5 SN data and by applying an ANN approach to align the BAO and SN data at the same redshifts. To obtain more comprehensive and robust results, we also incorporate the SDSS BAO and PantheonPlus SN data, and perform cross-checks using four pairwise independent combinations of the BAO and SN datasets.

This work is organized as follows. In Sect.~\ref{sec2}, we describe the BAO and SN data used in our analysis and outline the methodology for testing the CDDR. In Sect.~\ref{sec3}, we report the constraint results and make some relevant discussions. The conclusion is given in Sect.~\ref{sec4}.

\section{Methodology and data}\label{sec2}

We employ the function $\eta(z)$ to probe potential violations of the CDDR at various redshifts by comparing the luminosity distance from SN with the angular diameter distance derived from the BAO measurements. The definition of $\eta(z)$ is given by
\begin{equation}\label{eq1}
\eta(z) = \frac{D_{\rm L}}{D_{\rm A}(1+z)^2}.
\end{equation}
Any departure of $\eta(z)\neq1$ at a given redshift signifies an inconsistency between the CDDR and the cosmological data. We parameterize $\eta(z)$ using three forms: (i) $P_1$: $\eta(z) = 1 + \eta_0\,z$; (ii) $P_2$: $\eta(z) = 1 + \eta_0\,z/(1+z)$; (iii) $P_3$: $\eta(z) = 1 + \eta_0\,\ln(1+z)$. The first function represents the first-order term of a Taylor series expansion, but it exhibits undesirable behavior at high redshift. In contrast, the second function performs well even at high redshift, evolving more gradually than the first. The third function is expressed in a more complex logarithmic form. The reason for considering these three typical parameterizations is that they are widely used in many studies (see, e.g., Refs.~\cite{Liu:2021fka,Xu:2022zlm,Qi:2024acx,Yang:2024icv}), facilitating a comparison with our results. Furthermore, it is important to highlight that the motivation for considering multiple parameterizations stems from previous studies, which show that in any expansion of the form $\eta(z) = a + b f(z)$, the associated uncertainties are highly sensitive to the choice of the function $f(z)$ (see, e.g., Ref.~\cite{Colgain:2021pmf}). This inherent arbitrariness in a single parameterization is best addressed by considering multiple parameterizations, thus enabling a more robust and unbiased analysis.

The observed value $\eta_{\rm obs}(z)$ is computed using Eq.~\eqref{eq1}, and according to the error propagation formula, its uncertainty is given by
\begin{equation}\label{eq2}
\sigma^2_{\eta_{\rm obs}} 
= \eta_{\rm obs}^2 
  \left[
    \biggl(\frac{\sigma_{D_{\rm A}(z)}}{D_{\rm A}(z)}\biggr)^2
    +
    \biggl(\frac{\sigma_{D_{\rm L}(z)}}{D_{\rm L}(z)}\biggr)^2
  \right].
\end{equation}
Finally, the best-fit for the parameter $\eta_0$ is evaluated through
\begin{equation}\label{chi}
\chi^{2}(\eta_0)
= \sum_{i=1}^{N}
    \frac{\bigl[\eta(z_i) - \eta_{\rm obs,\,i}(z_i)\bigr]^{2}}
         {\sigma^2_{\eta_{\rm obs,\,i}}}\,.
\end{equation}

\begin{table*}[htbp]
\setlength\tabcolsep{15pt}
\renewcommand{\arraystretch}{1.0}
\centering
\caption{The BAO samples utilized in this work including DESI DR2 data and SDSS data.}
\label{tab:BAO_data}
\setlength\tabcolsep{15pt}{
\renewcommand\arraystretch{1.3}
\begin{tabular}{lccc}
\hline
\hline

Tracer & ${z_{\mathrm{eff}}}$ & ${D_{\mathrm{M}}/r_{\mathrm{d}}}$ & References \\

\midrule[0.4pt]
\multicolumn{3}{l}{\textbf{DESI}}\\
DR2 LRG1             & 0.51  & 13.59 $\pm$ 0.17 & \cite{DESI:2025zgx}\\
DR2 LRG2             & 0.71  & 17.35 $\pm$ 0.18 & \cite{DESI:2025zgx}\\
DR2 LRG3 + ELG1      & 0.93  & 21.58 $\pm$ 0.15 & \cite{DESI:2025zgx}\\
DR2 ELG2             & 1.32  & 27.60 $\pm$ 0.32 & \cite{DESI:2025zgx}\\
DR2 QSO              & 1.48  & 30.51 $\pm$ 0.76 & \cite{DESI:2025zgx}\\
\hline
\multicolumn{3}{l}{\textbf{SDSS}}\\
BOSS LRG1           & 0.38  & 10.27 $\pm$ 0.15 & \cite{Gil-Marin:2016wya}\\
BOSS LRG2           & 0.51  & 13.38 $\pm$ 0.18 & \cite{Gil-Marin:2016wya}\\
eBOSS LRG           & 0.70  & 17.65 $\pm$ 0.30 & \cite{eBOSS:2020lta}\\
eBOSS ELG           & 0.85  & 19.50 $\pm$ 1.00 & \cite{eBOSS:2020gbb}\\
eBOSS QSO           & 1.48  & 30.21 $\pm$ 0.79 & \cite{eBOSS:2020gbb}\\
\hline
\hline
\end{tabular} 
}
\end{table*}

The data of $D_{\rm L}$ we use in this work are obtained from following two SN datasets:
\begin{itemize}
\item \textbf{\texttt{PantheonPlus}:} The PantheonPlus comprises 1550 spectroscopically confirmed supernovae from 18 different surveys, covering the redshift range $0.01 < z < 2.26$ \cite{Brout:2022vxf}.\footnote{\url{https://github.com/PantheonPlusSH0ES/DataRelease}.}
\item \textbf{\texttt{DESY5}:}  The DESY5 sample comprises 1635 photometrically classified supernovae from the released part of the full 5-year data of the Dark Energy Survey collaboration (with a redshift range $0.1 < z < 1.3$), complemented by 194 low-redshift supernovae from the CfA3~\cite{Hicken:2009df}, CfA4~\cite{Hicken:2012zr}, CSP~\cite{Krisciunas:2017yoe}, and Foundation~\cite{Foley:2017zdq} samples (with redshifts in the range $0.025 < z < 0.1$), for a total of 1829 supernovae \cite{DES:2024jxu}.\footnote{\url{https://github.com/des-science/DES-SN5YR}.}  
\end{itemize}

The SN distance modulus $\mu$ is given by $\mu = m_{\rm B} - M_{\rm B}$, where $m_{\rm B}$ represents the observed peak apparent magnitude in the rest-frame $B$ band, $M_{\rm B}$ is the absolute magnitude. The luminosity distance $D_{\rm L}(z)$ connects to $\mu(z)$ through
\begin{equation}
\mu(z)=5\log_{10}\biggl[\frac{D_{\rm L}(z)}{\mathrm{Mpc}}\biggr]+25,
\end{equation}
with the corresponding uncertainty in $D_{\rm L}$ derived by standard error propagation
\begin{equation}\label{eq5}
\sigma_{D_{\rm L}} = \frac{D_{\rm L}\ln 10}{5}\sigma_{\mu}.
\end{equation}

\begin{figure*}
\centering
\includegraphics[width=0.49\textwidth]{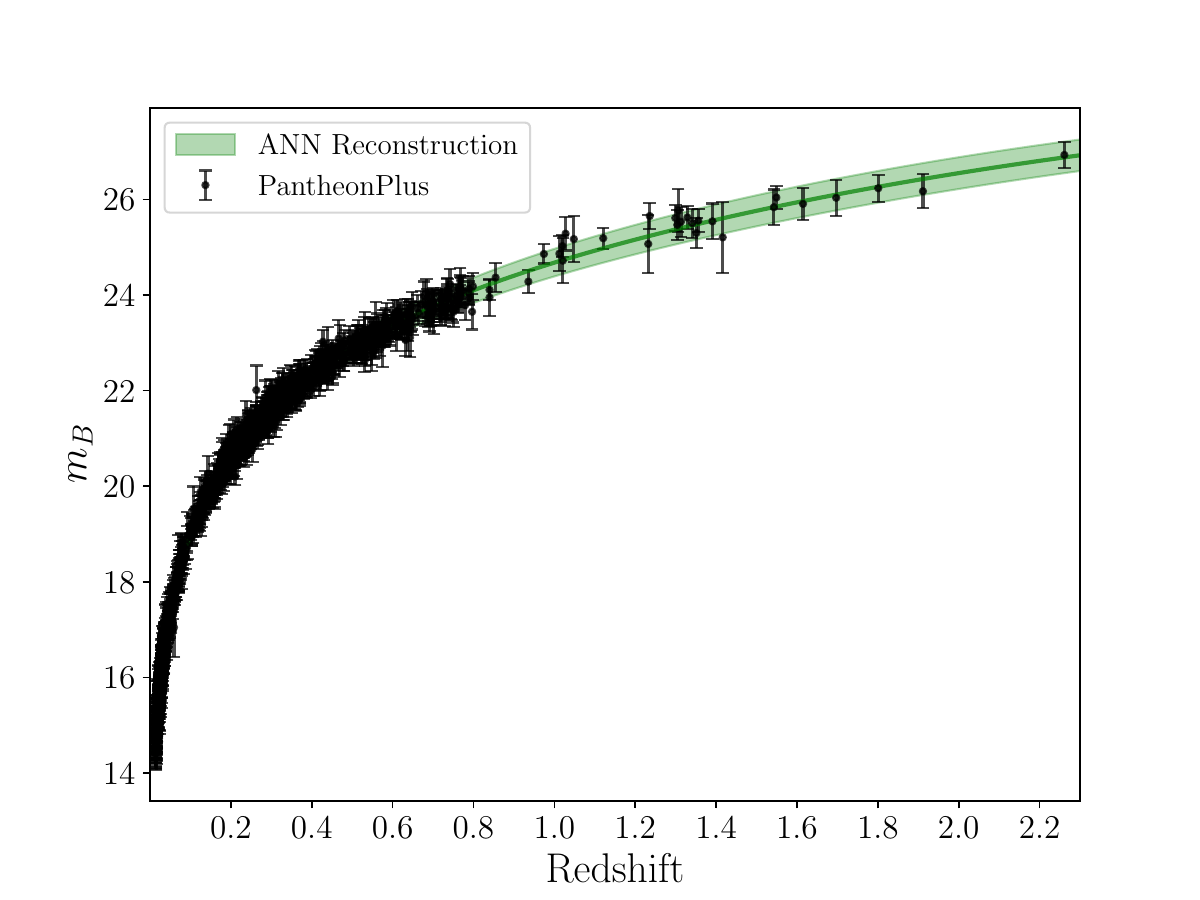} \hspace{2pt}
\includegraphics[width=0.49\textwidth]{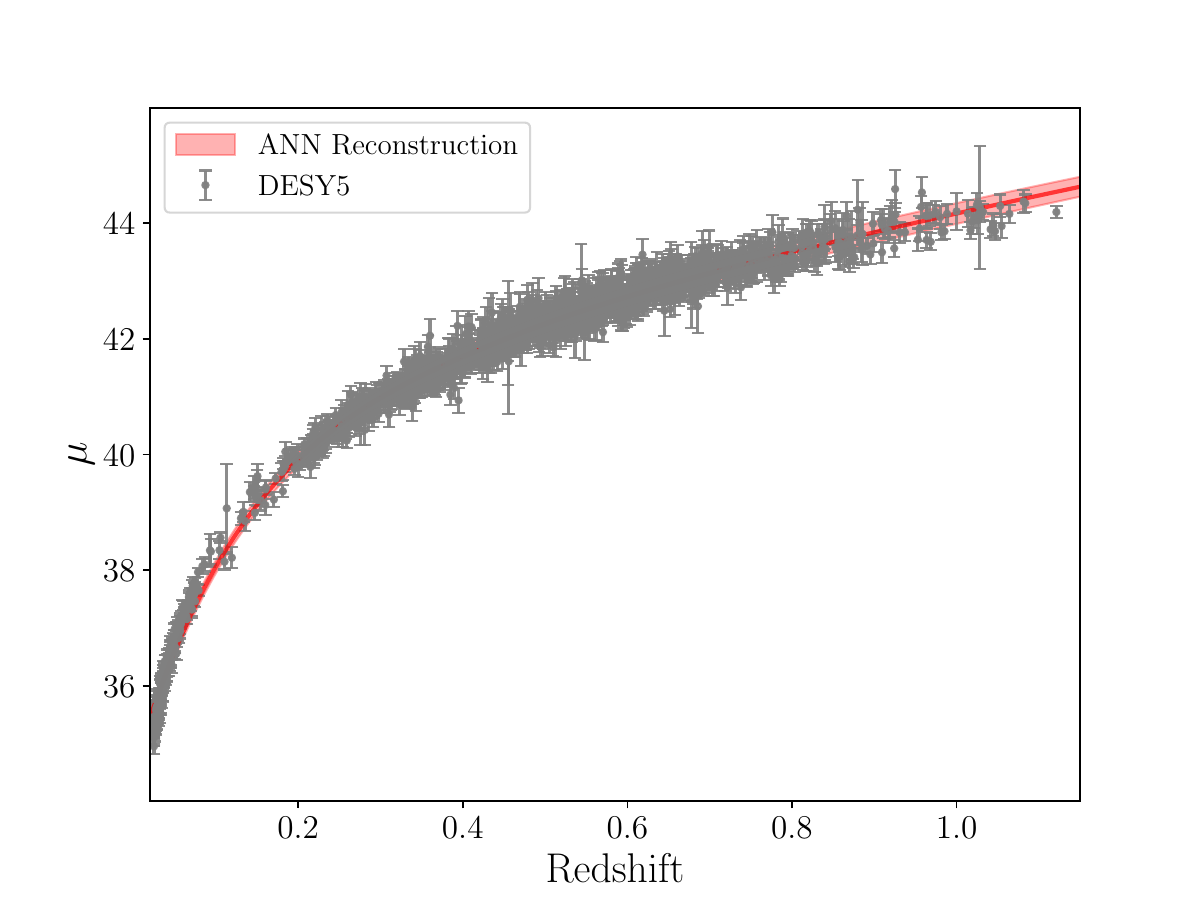}
\caption{\label{fig1} The ANN reconstruction for the SN data. \emph{Left panel}: The ANN-reconstructed $m_{\rm B}$ for the PantheonPlus over the redshift range $0.01\leq z\leq2.26$. The light green shaded band denotes the $1\sigma$ confidence interval of the ANN reconstruction, while the black points with error bars correspond to the PantheonPlus data points. \emph{Right panel}: The ANN-reconstructed $\mu$ for the DESY5 over the redshift range $0.025\leq z\leq1.3$. The light red shaded band denotes the $1\sigma$ confidence interval of the ANN reconstruction, and the light grey points with error bars represent the DESY5 data points.}
\end{figure*}

The distance measurements of SN crucially rely on the calibration of $M_{\rm B}$. Using the inverse distance ladder, the CMB constraint on the sound horizon predicts $M_{\rm B} \sim -19.4$ mag, while the estimate from SH0ES corresponds to $M_{\rm B} \sim -19.2$ mag. In our analysis, for PantheonPlus, we first perform fits treating $M_{\rm B}$ as a free parameter to test CDDR. Additionally, it is important to further study the impacts of different prior values of $M_{\rm B}$ on the CDDR test. Hence, we adopt two specific Gaussian priors of $M_{\rm B}$ derived from different observational datasets within various redshift ranges: (i) $M_{\rm B}^{\rm D20} = -19.230 \pm 0.040\,\mathrm{mag}$, from low‑$z$ SN ($0.023<z<0.15$) in $\Lambda$CDM via de‑marginalization of the SH0ES result~\cite{Camarena:2019rmj,Reid:2019tiq}; (ii) $M_{\rm B}^{\rm B23} = -19.396 \pm 0.016\,\mathrm{mag}$, from combined SN and BAO data~\cite{Dinda:2022jih}. Including the observational error in $M_{\rm B}$, the uncertainty in the distance modulus is $\sigma_{\mu} =\sqrt{\sigma_{M_{\rm B}}^2 + \sigma_{m_{\rm B}}^2}\,$. Conversely, the DESY5 compilation provides $\mu$ and $\sigma_{\mu}$ directly, as $M_{\rm B}$ is already normalized within the calibration. Therefore, no prior on $M_{\rm B}$ is imposed for DESY5 analysis in this work.

For the measurement of $D_{\rm A}(z)$, we utilize the transverse comoving distance $D_{\mathrm{M}}(z)/r_{\mathrm{d}}$ derived from the BAO observations of the DESI DR2 \cite{DESI:2025zgx} and SDSS \cite{Gil-Marin:2016wya,eBOSS:2020gbb,eBOSS:2020lta}. The specific BAO measurements are listed in Table~\ref{tab:BAO_data}. The angular diameter distance $D_{\rm A}(z)$ is obtained through
\begin{equation}
    D_{\rm A}(z) = \frac{D_{\rm M}(z)}{1+z}.
\end{equation}
In this work, we adopt the sound horizon $r_{\mathrm{d}} = 147.78~\mathrm{Mpc}$ corresponding to the Planck 2018 result \cite{Planck:2018vyg}. Note that we do not consider the uncertainty in $r_{\rm d}$, since it is much smaller than the uncertainties in $D_{\rm L}(z)$ and $D_{\rm A}(z)$ and is therefore negligible. Additionally, some DESI and SDSS BAO measurements are excluded from this analysis because their redshifts lie beyond the range probed by the SN samples. Specifically, when using the PantheonPlus SN sample, we omit BAO measurements from the Lyman-$\alpha$ forest in DESI DR2 and SDSS eBOSS. When using the DESY5 SN sample, we omit the DESI DR2 ELG2, QSO, and Lyman-$\alpha$ forest measurements, as well as the SDSS eBOSS QSO and Lyman-$\alpha$ forest data points.

\begin{table*}[htbp]
\centering
\caption{\label{table2}Fitting results ($1\sigma$ confidence level) in the $P_1$, $P_2$, and $P_3$ models from the BAO and SN data.}
\renewcommand{\arraystretch}{2.2}
\setlength{\tabcolsep}{9pt}
\scriptsize % 
\setlength{\tabcolsep}{6pt} % 
\begin{tabular*}{\textwidth}{@{\extracolsep{\fill}}>{\rule{0pt}{2.5ex}}lcccccc}

\hline\hline
\textbf{Data}\rule{0pt}{3ex} 
    & \multicolumn{2}{c}{\textbf{$P_1$} model} 
    & \multicolumn{2}{c}{\textbf{$P_2$} model} 
    & \multicolumn{2}{c}{\textbf{$P_3$} model} \\
& $\eta_0$\rule{0pt}{1.5ex} & $M_{\rm B}$ 
& $\eta_0$ & $M_{\rm B}$ 
& $\eta_0$ & $M_{\rm B}$ \\

\hline
SDSS + PantheonPlus  
& $0.040^{+0.130}_{-0.190}$  & $-19.510^{+0.290}_{-0.230}$  
& $0.290^{+0.360}_{-0.810}$   & $-19.630^{+0.580}_{-0.400}$  
& $0.100^{+0.220}_{-0.390}$  & $-19.540^{+0.400}_{-0.300}$ \\
DESI + PantheonPlus  
& $0.060^{+0.120}_{-0.210}$  & $-19.560^{+0.420}_{-0.290}$  
& $0.480^{+0.440}_{-0.990}$   & $-19.810^{+0.790}_{-0.570}$  
& $0.170^{+0.220}_{-0.480}$  & $-19.650^{+0.570}_{-0.390}$ \\

SDSS + PantheonPlus $\times M_{\rm B}^{\rm D20}$  
& $-0.120\pm 0.067$ & $-19.230\pm 0.040$  
& $-0.240\pm 0.130$   & $-19.230\pm 0.040$  
& $-0.178\pm 0.093$   & $-19.230\pm 0.040$ \\
DESI + PantheonPlus $\times  M_{\rm B}^{\rm D20}$  
& $-0.098\pm 0.052$ & $-19.230\pm 0.040$  
& $-0.220\pm 0.113$   & $-19.230\pm 0.040$  
& $-0.162\pm 0.068$   & $-19.230\pm 0.040$ \\

SDSS + PantheonPlus $\times M_{\rm B}^{\rm B23}$ 
& $-0.040\pm 0.071$ & $-19.396\pm 0.016$  
& $-0.052\pm 0.131$   & $-19.396\pm 0.016$  
& $-0.050\pm 0.102$ & $-19.396\pm 0.016$ \\
DESI + PantheonPlus $\times M_{\rm B}^{\rm B23}$ 
& $-0.035\pm 0.056$ & $-19.396\pm 0.016$  
& $-0.070\pm 0.124$   & $-19.396\pm 0.016$  
& $-0.043\pm 0.088$ & $-19.396\pm 0.016$ \\

SDSS + DESY5  
& $-0.035\pm 0.074$ & $-$  
& $-0.080\pm 0.120$   & $-$  
& $-0.059\pm 0.093$ & $-$ \\
DESI + DESY5  
& $-0.052\pm 0.066$ & $-$  
& $-0.090\pm 0.110$   & $-$  
& $-0.068\pm 0.085$ & $-$ \\
\hline\hline
\end{tabular*}
\end{table*}

In order to match the BAO and SN data at a given redshift, we reconstruct $m_{\rm B}$ or $\mu$ of SN using a nonparametric ANN method implemented with the \texttt{REFANN} Python package \cite{Wang:2019vxv}. This ANN method is fully data-driven, allowing us to reconstruct a function from any type of data without assuming a specific parameterization of the function. An ANN typically consists of three main layers: the input layer, the hidden layer, and the output layer. The input layer consists of the features of the dataset, with each neuron representing a feature or variable. Since our goal is to reconstruct the functional relationship from the real data, we use all available data as the training input, without retaining a separate testing set, which is consistent with many similar studies (see, e.g., Refs.~\cite{Qi:2024acx,Yang:2024icv}). The neurons in the hidden layer receive outputs from the previous layer and transmit them to the neurons in the next layer. The hidden layers in an ANN are responsible for learning and data processing, and can consist of one or more layers. In our analysis, a single hidden layer connects the supernova redshift with the corresponding $m_{\rm B}$ or $\mu$ values, with a total of 4096 neurons in the hidden layer. This choice of the number of neurons is consistent with our previous work, providing a reasonable balance between accuracy and computational efficiency \cite{Qi:2023oxv,Qi:2024acx}. The output layer is primarily responsible for generating the final result. We use a nonlinear activation function (exponential linear unit), with the remaining hyperparameters set to their default values as specified in Ref.~\cite{Wang:2019vxv}, and the mean absolute error is used as the loss function. The optimization method employed is gradient descent, which iteratively updates the loss value in the opposite direction of the current gradient to reduce the loss. The training process is performed over 30000 iterations to ensure the convergence of the loss function.

The ANN method is trained on the PantheonPlus and DESY5 datasets to yield a data-driven reconstruction of $m_{\rm B}$ or $\mu$. It is worth noting that for the SN data, we used only the diagonal elements of the covariance matrix to facilitate reconstruction using ANN. From this reconstruction we extract $D_{\rm L}(z)$ at the redshifts of BAO measurements and compare it with the angular diameter distance $D_{\rm A}(z)$ inferred from the same BAO data to test the CDDR. The ANN reconstruction $m_{\rm B}$ for PantheonPlus and $\mu$ for DESY5 in this work are shown in Fig.~\ref{fig1}. Furthermore, we construct the SN Hubble diagram and compute the distance duality ratio $\eta(z)$, together with their uncertainties at the corresponding BAO redshifts using Eqs.~(\ref{eq1}), (\ref{eq2}), and (\ref{eq5}). The detailed results and discussion are presented in Appendix~\ref{appendixA}.

This nonparametric framework does not assume any predefined functional form, providing a flexible approach for probing potential deviations from the standard cosmological model \cite{Wang:2020dbt,Liu:2021fka,Qi:2023oxv,Liu:2024yib,Abedin:2025yru}. The precision of the reconstructed $m_{\rm B}$ or $\mu$ and the width of their confidence intervals depend on hyperparameter choices, such as network architecture and training strategy, which influence the expressive power and generalization capability of ANN. Nonetheless, the ANN‑based reconstruction remains a powerful, model‑independent tool for evaluating the CDDR using BAO and SN data.

We perform Bayesian analysis using the publicly available package {\tt Cobaya}\footnote{\url{https://github.com/CobayaSampler/cobaya}.} \cite{Torrado:2020dgo} and assess the convergence of the Markov Chain Monte Carlo (MCMC) \cite{Lewis:2002ah,Lewis:2013hha} chains using the Gelman-Rubin statistics quantity $R - 1 < 0.01$ \cite{Gelman:1992zz}. The MCMC chains are analyzed using {\tt GetDist}\footnote{\url{https://github.com/cmbant/getdist}.} \cite{Lewis:2019xzd}.

\section{Results and discussions}\label{sec3}

\begin{figure*}[htbp]
\centering
\includegraphics[width=5.5cm]{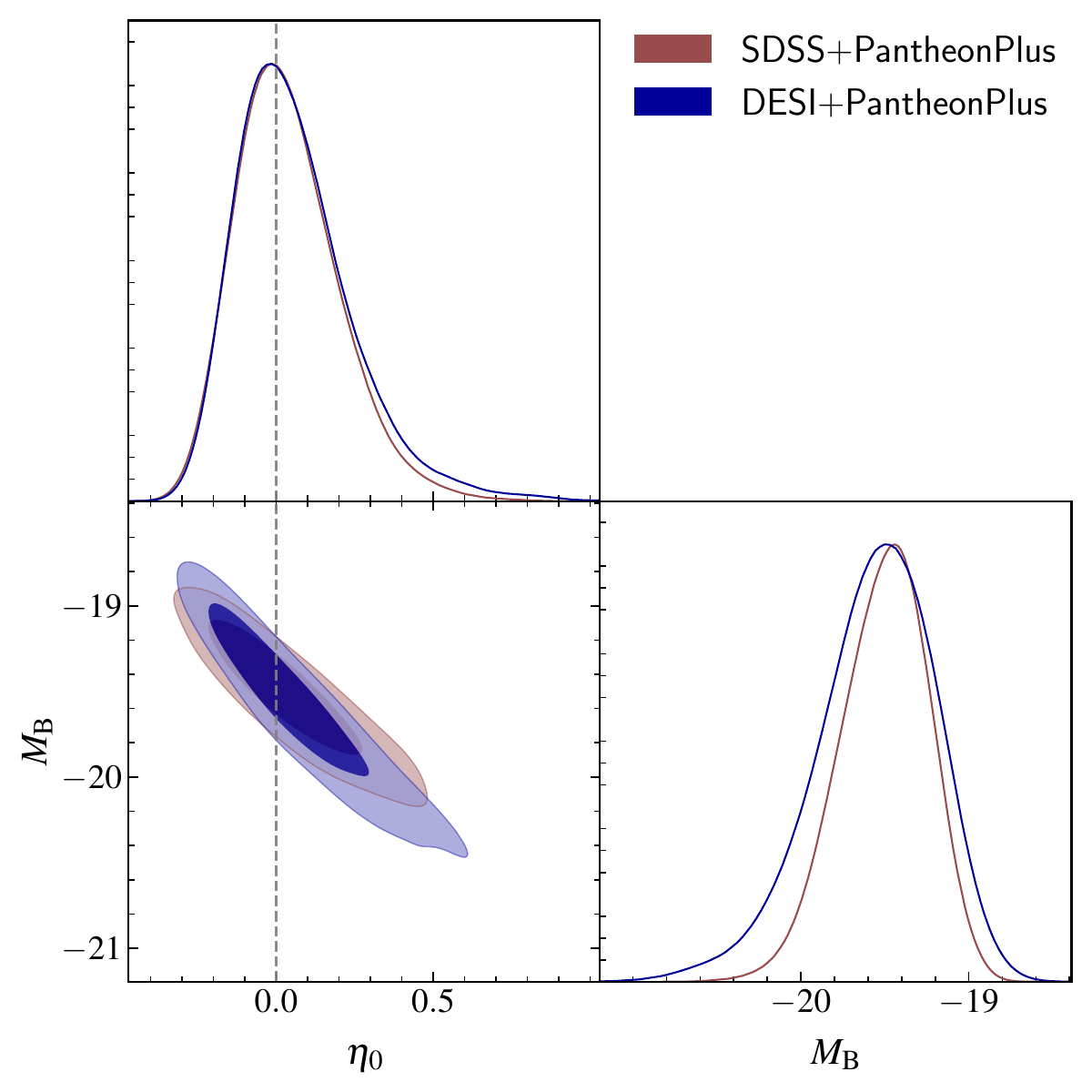}
\includegraphics[width=5.5cm]{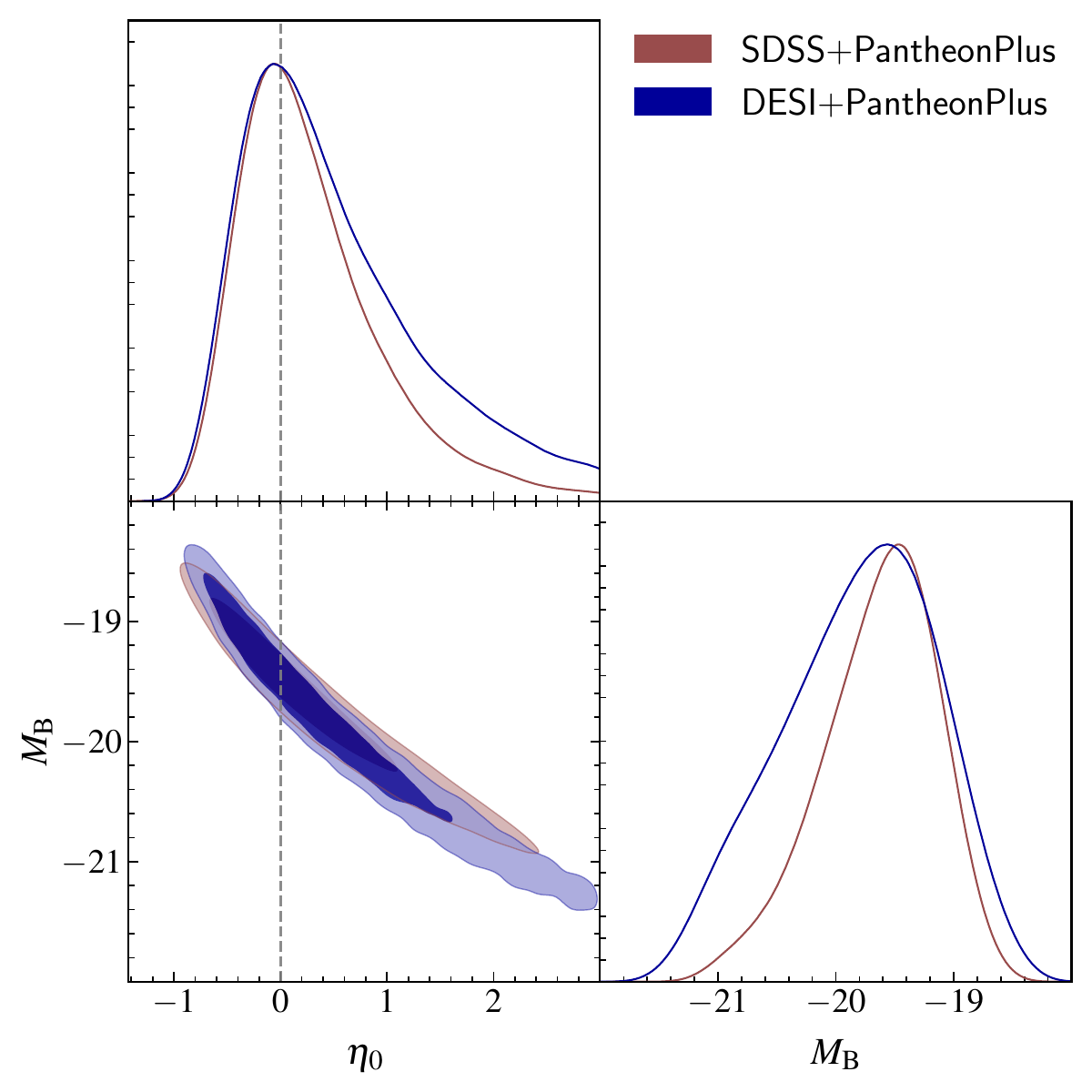}
\includegraphics[width=5.5cm]{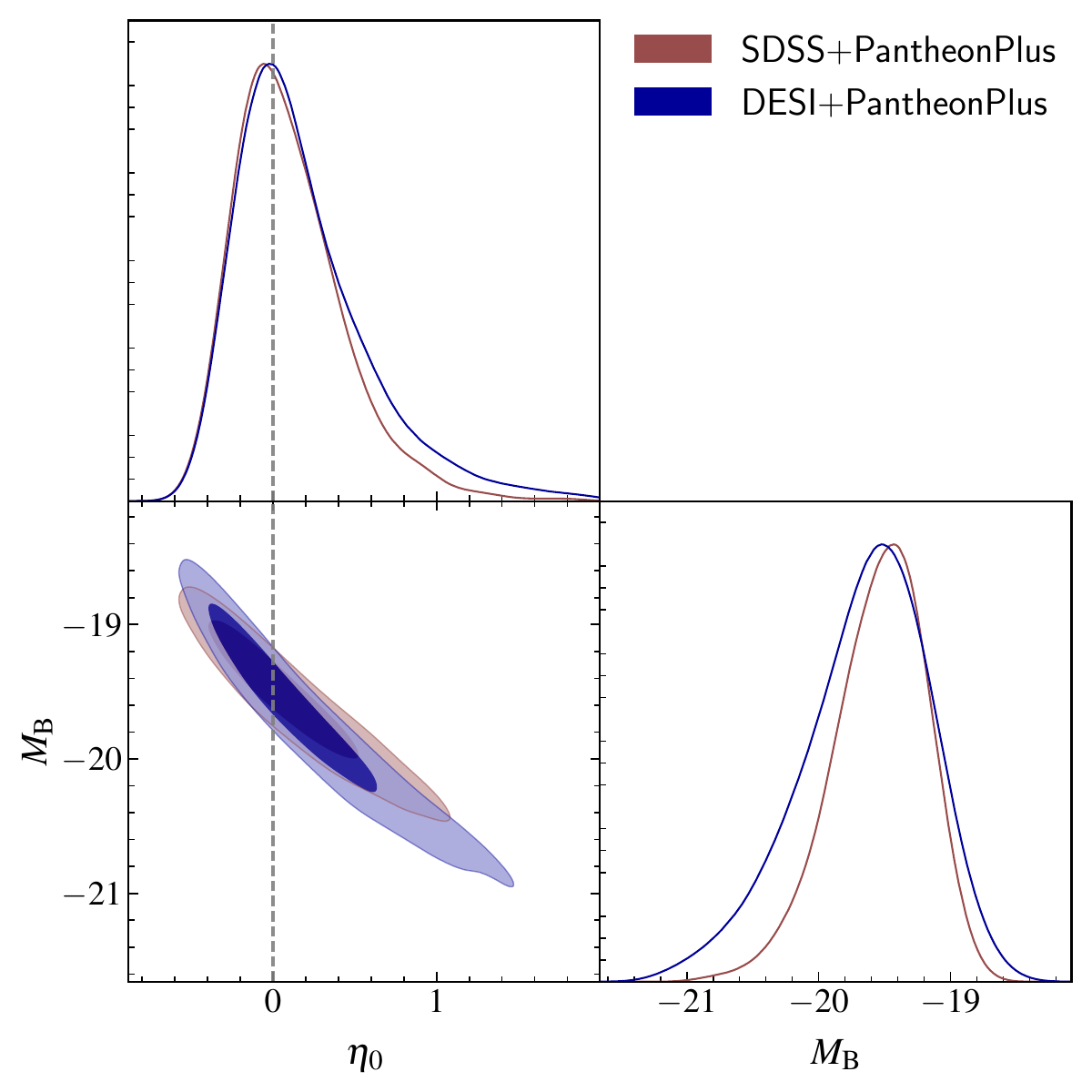}
\caption{\label{fig2}The triangular plots on $\eta_0$ and $M_{\rm B}$ for the $P_1$ (left panel), $P_2$ (middle panel), and $P_3$ (right panel) models, using SDSS, DESI, and PantheonPlus data.}
\end{figure*}

\begin{figure*}[htbp]
\centering
\includegraphics[width=5.4cm]{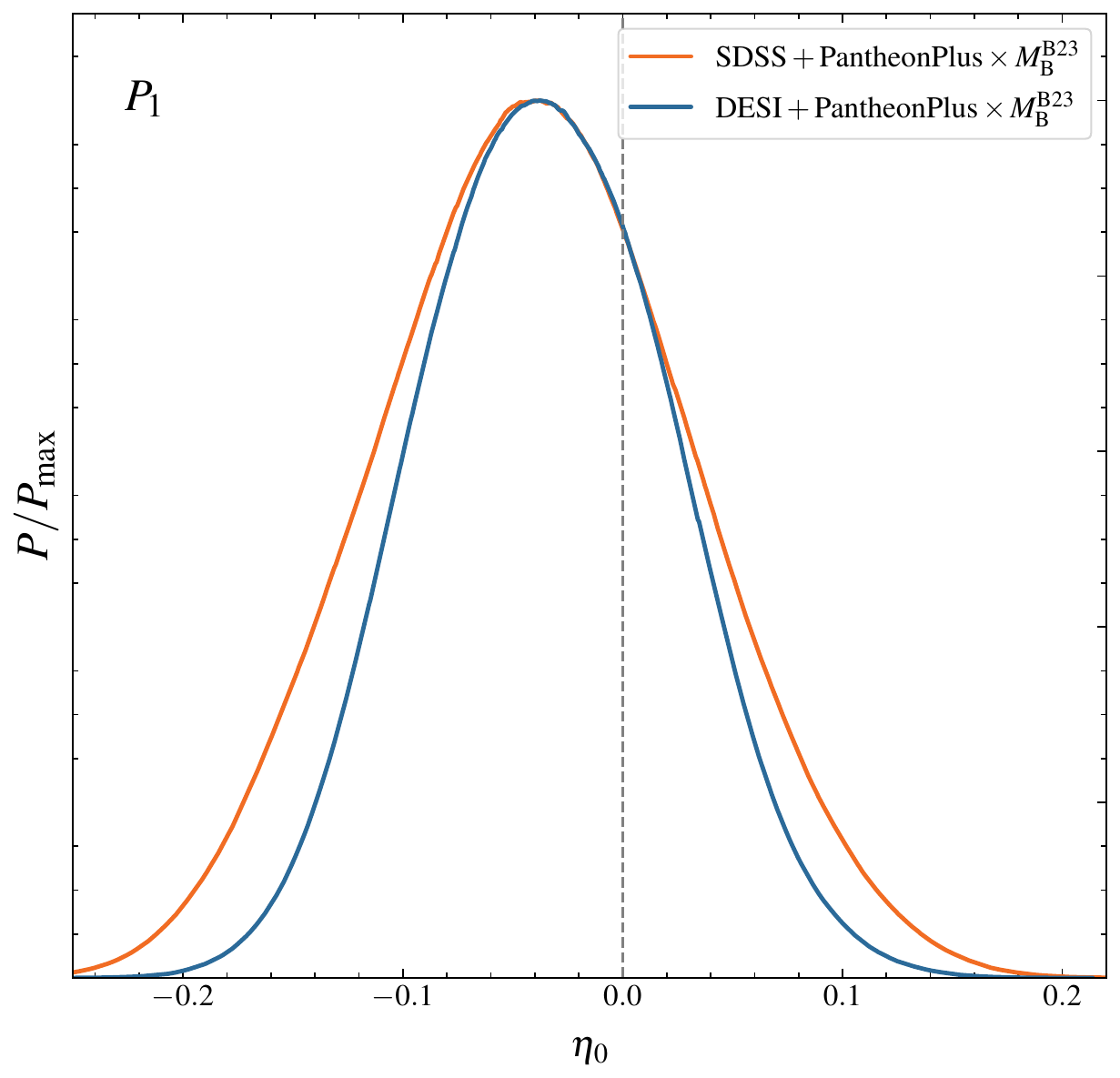}
\includegraphics[width=5.5cm]{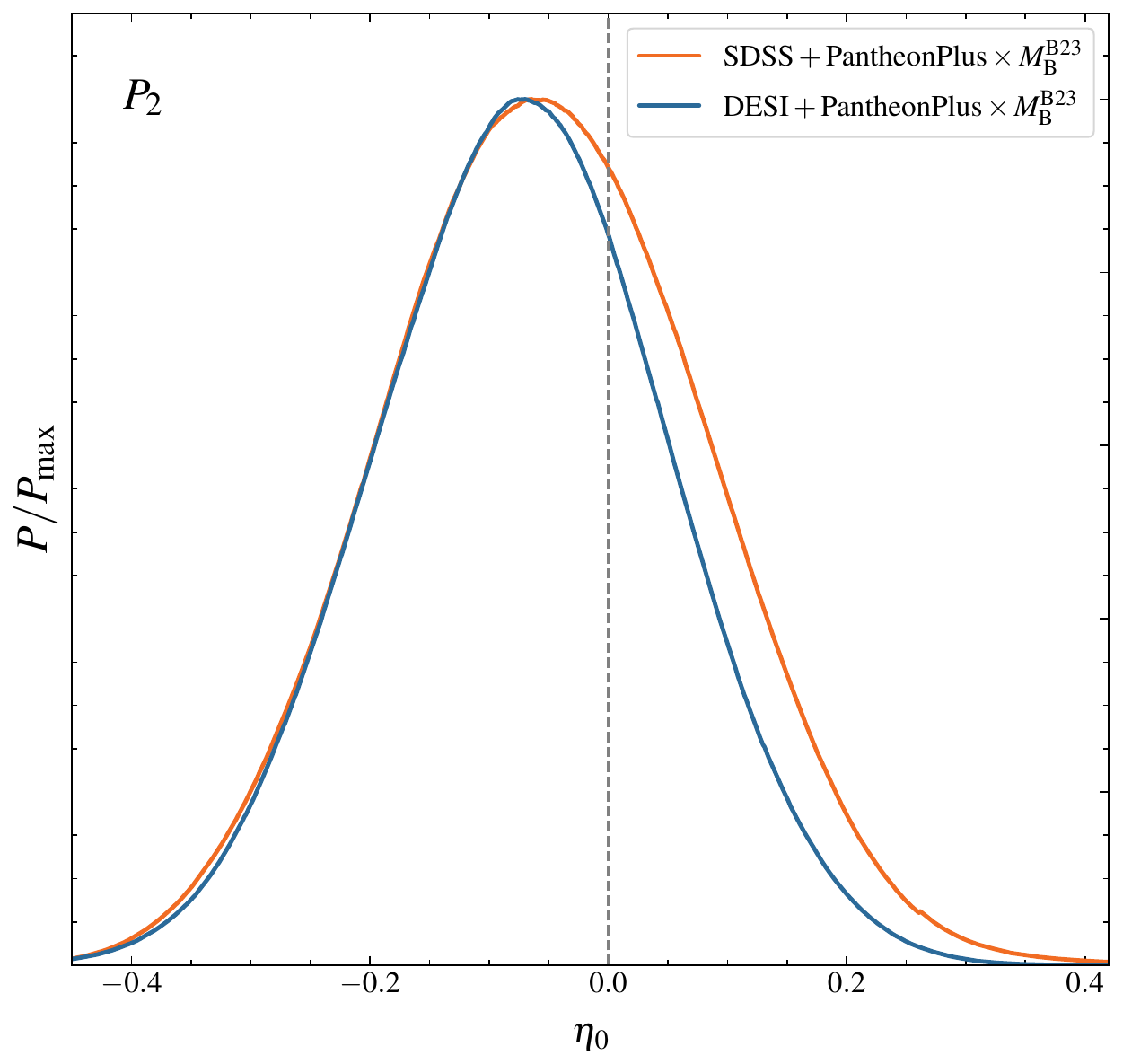}
\includegraphics[width=5.5cm]{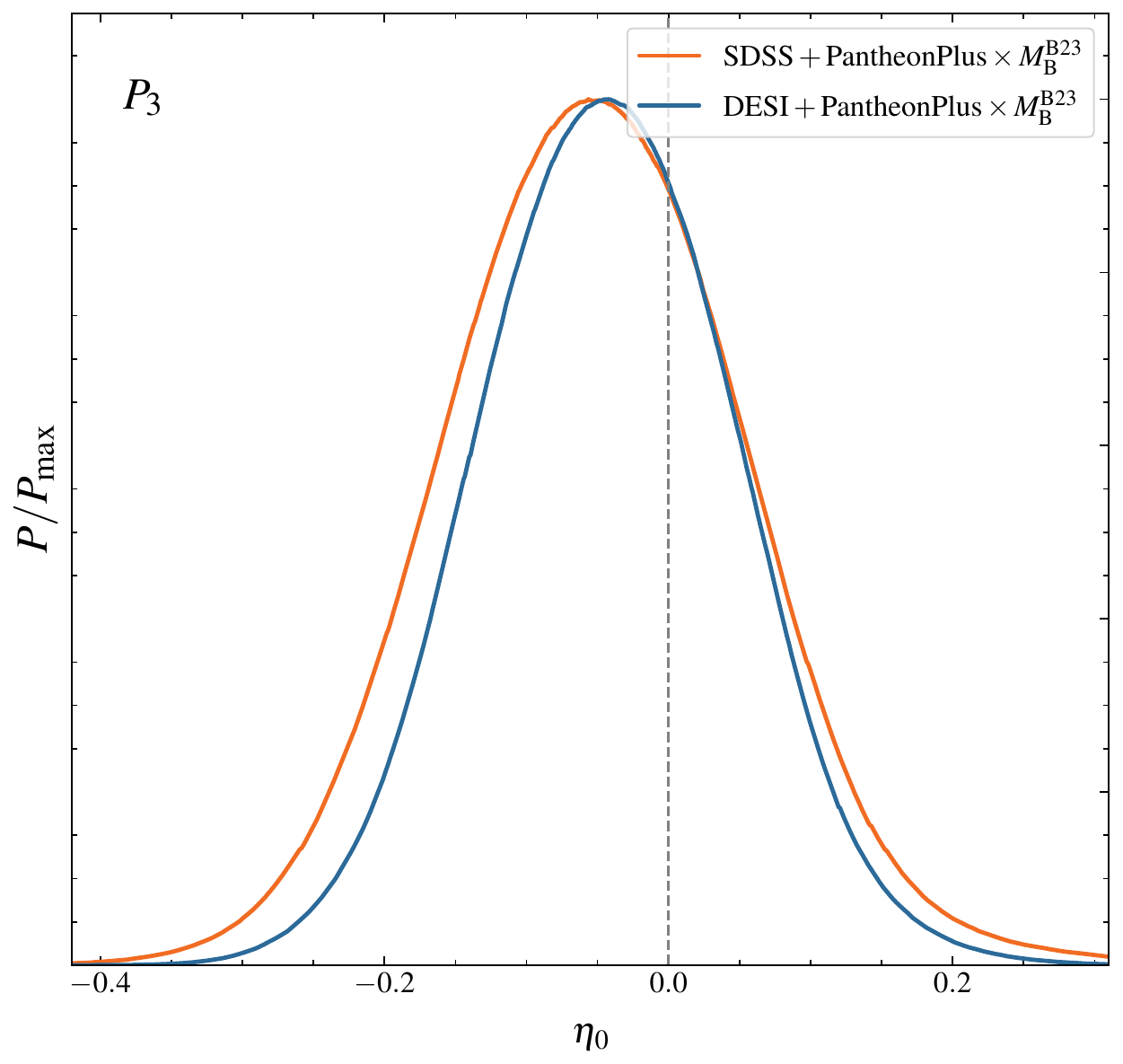}
\includegraphics[width=5.5cm]{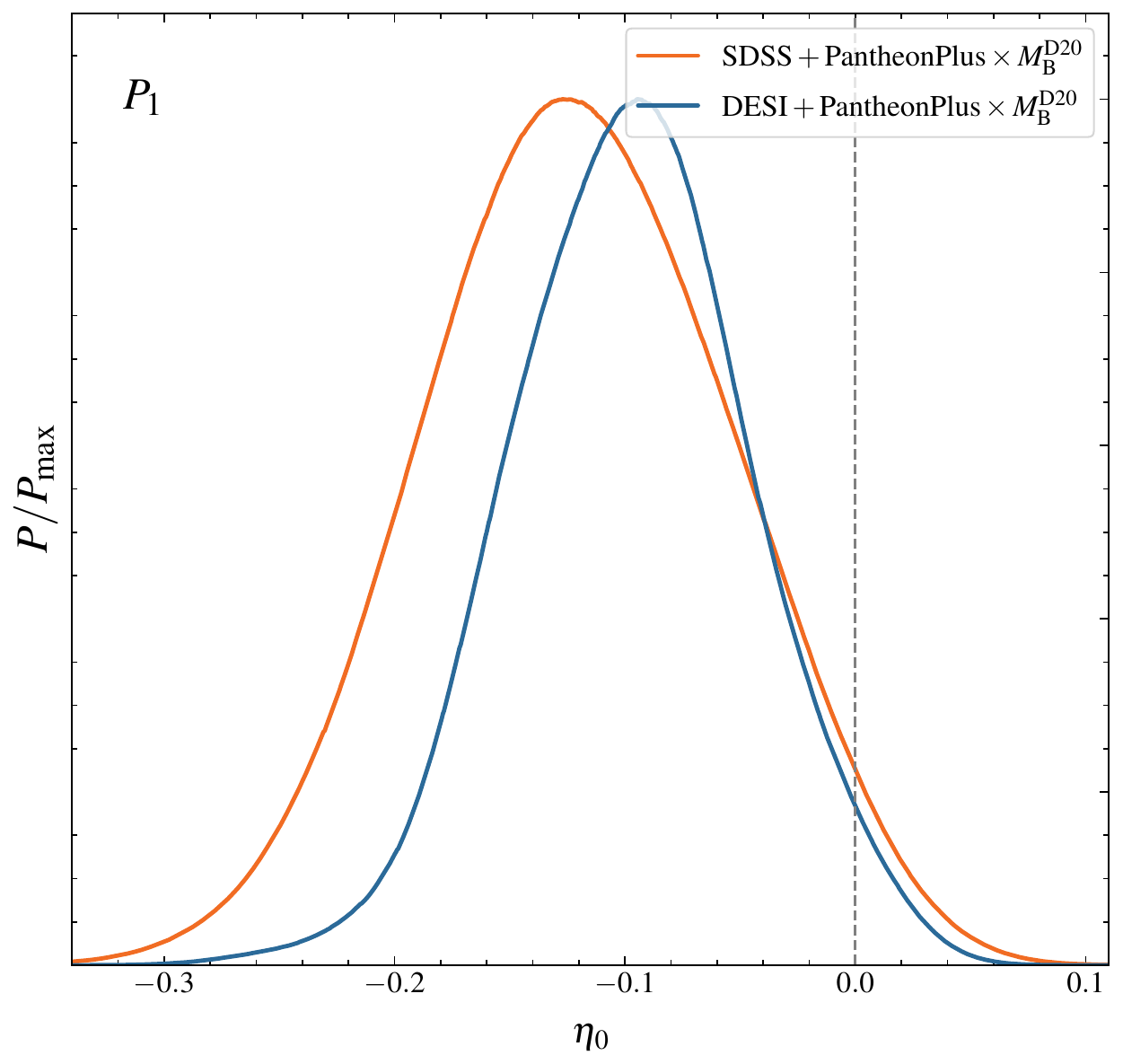}
\includegraphics[width=5.5cm]{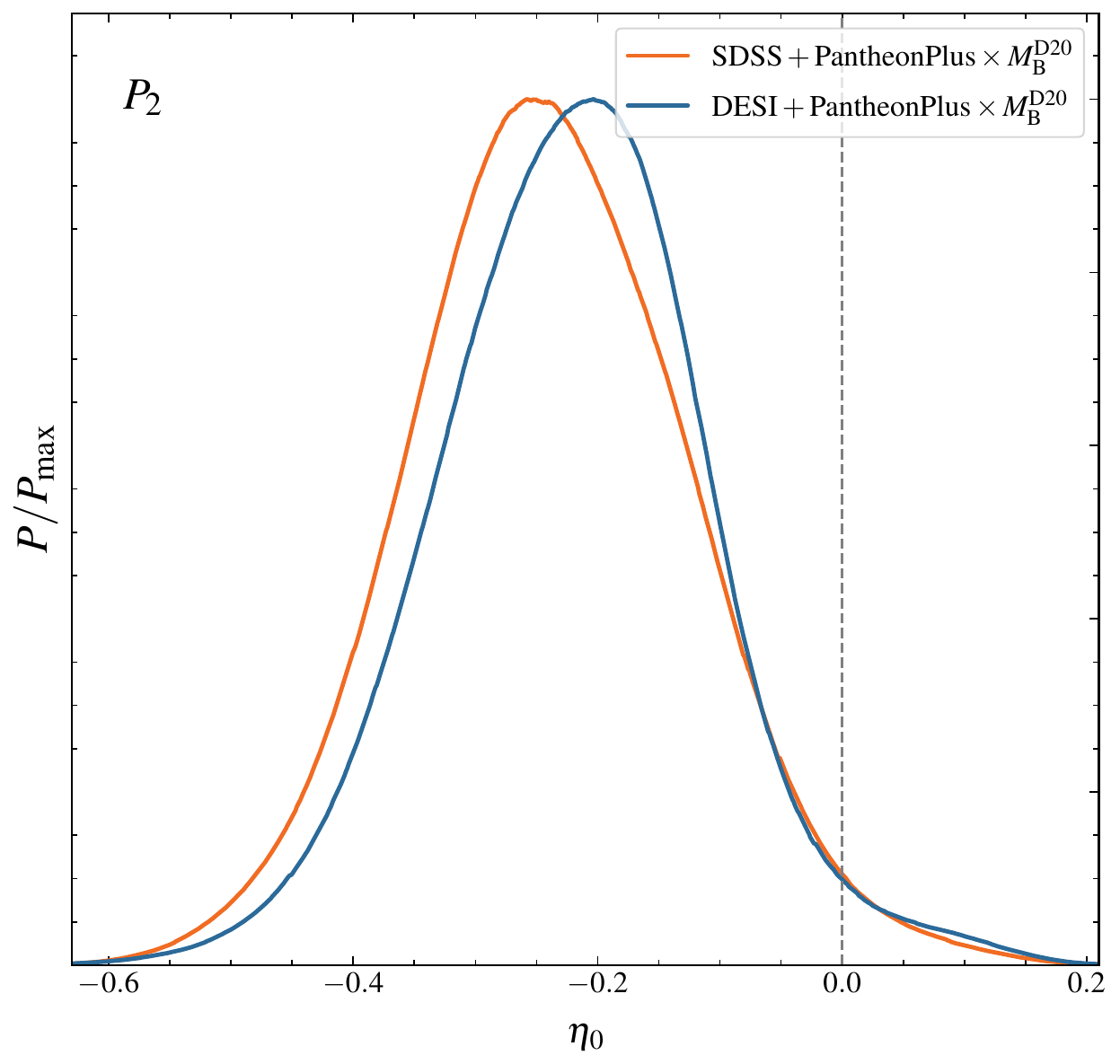}
\includegraphics[width=5.5cm]{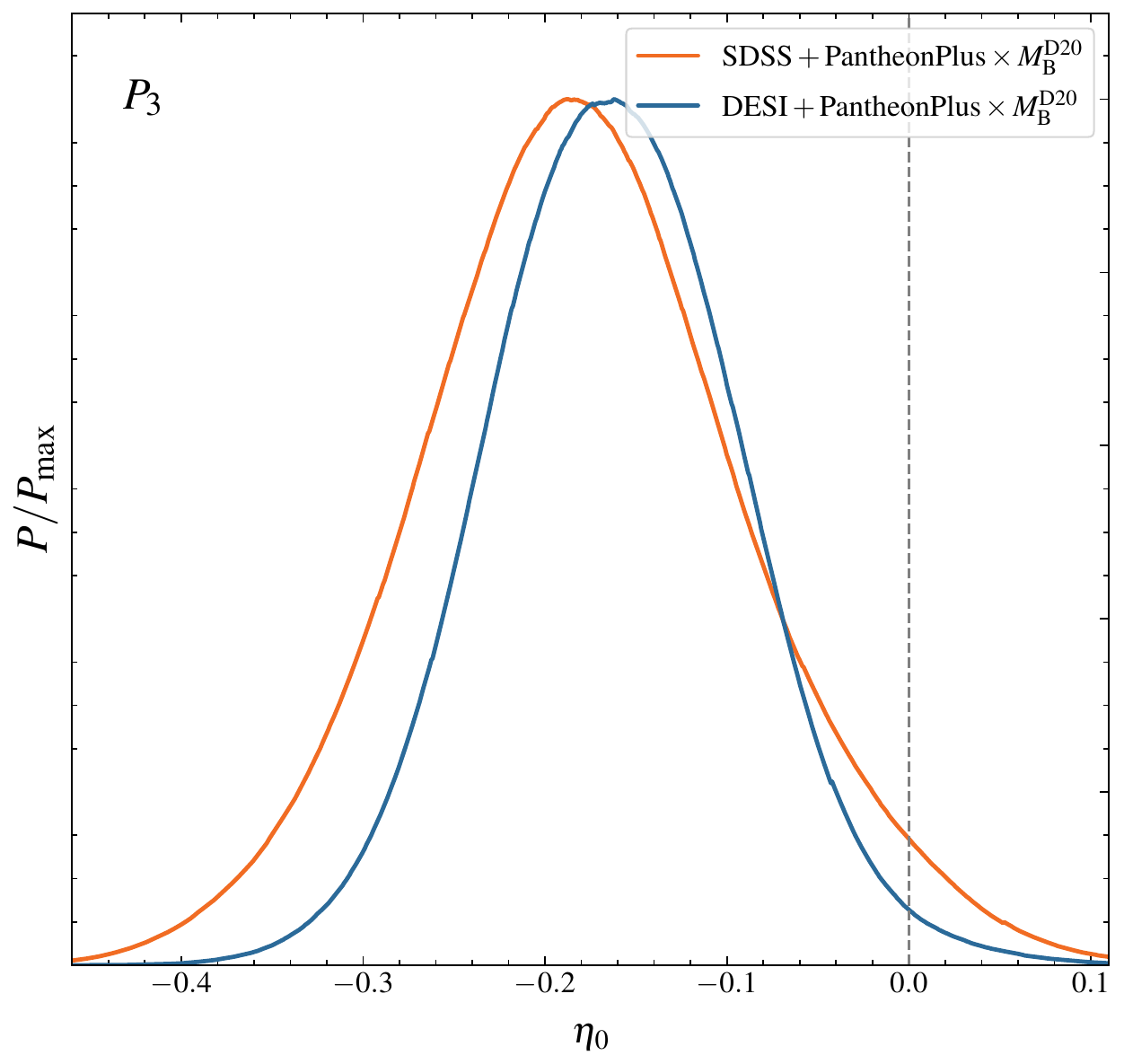}

\caption{\label{fig3} The 1D marginalized posterior distributions of the $\eta_0$ parameter for the $P_1$, $P_2$, and $P_3$ models of the CDDR with $M_{\rm B}$ fixed, using $M_{\rm B}^{\rm B23}$ (upper panel) and $M_{\rm B}^{\rm D20}$ (lower panel), based on SDSS, DESI, and PantheonPlus data.}
\end{figure*}

\begin{figure*}[!htp]
\includegraphics[scale=0.4]{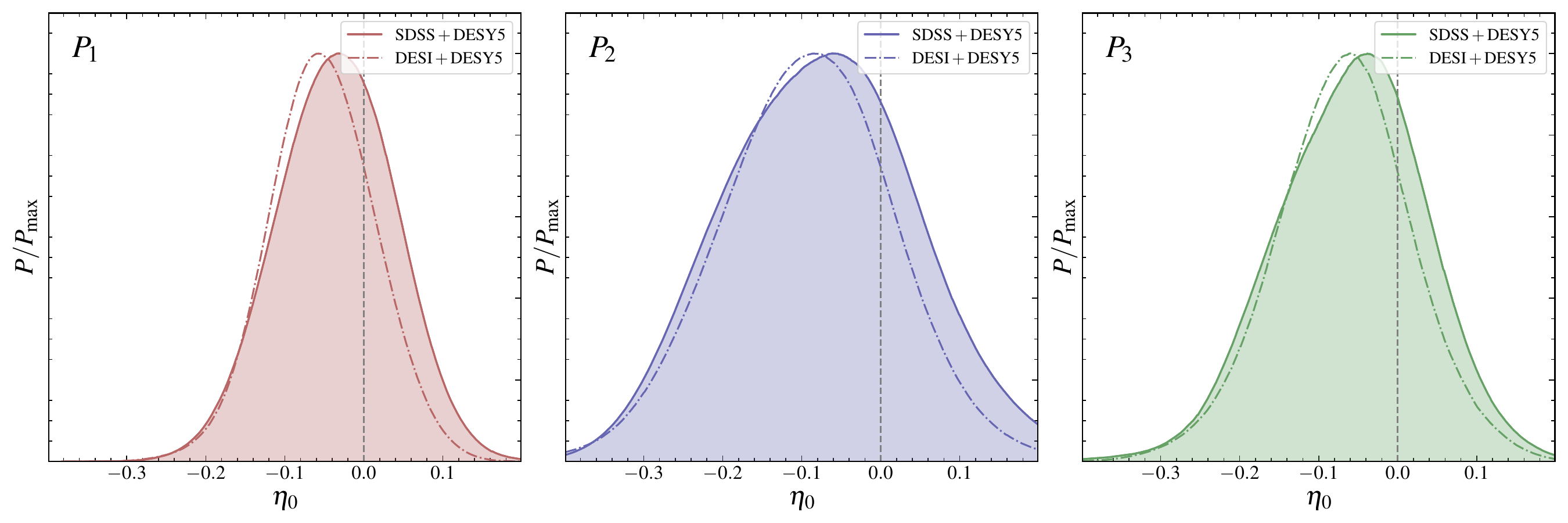}
\centering
\caption{\label{fig4}The 1D marginalized posterior distributions of the $\eta_0$ parameter for the $P_1$, $P_2$, and $P_3$ models of the CDDR, using SDSS, DESI, and DESY5 data.}
\end{figure*}

In this section, we present the constraints on $\eta_0$ and $M_{\rm B}$ for the $P_1$, $P_2$, and $P_3$ parameterizations of $\eta(z)$, using the SDSS, DESI, PantheonPlus, and DESY5 data. The marginalized results with $1\sigma$ errors are summarized in Table~\ref{table2}. The triangular plots, along with the one-dimensional (1D) marginalized distributions of $\eta_0$ and $M_{\rm B}$, based on various datasets and parameterizations, are shown in Figs.~\ref{fig2},~\ref{fig3}~and~\ref{fig4}.

In Fig.~\ref{fig2}, we show the triangular plot of the constraint results obtained by treating $M_{\rm B}$ as a free parameter, using SDSS, DESI, and PantheonPlus data. The constraints on $\eta_0$ are $0.060^{+0.120}_{-0.210}$ ($P_1$ model), $0.480^{+0.440}_{-0.990}$ ($P_2$ model), and $0.170^{+0.220}_{-0.480}$ ($P_3$ model), based on the DESI+PantheonPlus data. This indicates that in all parameterized models, there is no evidence for a violation of the CDDR. This result emphasizes the robustness of our CDDR analysis, and the consistency among the three parameterized models $P_1$, $P_2$, and $P_3$ also suggests that using multiple parameterizations of $\eta(z)$ can provide complementary insights into potential deviations from the CDDR. Similarly, the SDSS+PantheonPlus combination yields consistent results, indicating good agreement between the BAO measurements from SDSS and DESI DR2. The constraints on $M_{\rm B}$ using the DESI+PantheonPlus datasets are $-19.560^{+0.420}_{-0.290}$ ($P_1$ model), $-19.810^{+0.790}_{-0.570}$ ($P_2$ model), and $-19.650^{+0.570}_{-0.390}$ ($P_3$ model). The central values of $M_{\rm B}$ in all parameterized models are slightly lower than the commonly accepted value of approximately $-19.3$ mag. This slight shift may result from the inclusion of the possible deviation parameter $\eta_0$ in the CDDR analysis.

It is worth noting that there is a strong negative correlation between $\eta_0$ and $M_{\rm B}$, as shown in Fig.~\ref{fig2}. This correlation indicates that different $M_{\rm B}$ priors will have a significant impact on our detection of potential deviations from the CDDR. Therefore, we further adopt the $M_{\rm B}^{\rm D20}$ and $M_{\rm B}^{\rm B23}$ priors to assess the potential impact of $M_{\rm B}$ on deviations from the CDDR. The 1D marginalized posterior distributions of $\eta_0$ for $P_1$, $P_2$, and $P_3$ models are shown in Fig.~\ref{fig3}. When adopting the $M_{\rm B}^{\rm D20}$ prior with $M_{\rm B}=-19.230\pm0.040$ mag, $P_1$, $P_2$, and $P_3$ models yield $\eta_0=-0.098\pm0.052$, $\eta_0=-0.220\pm0.113$, and $\eta_0=-0.162\pm0.068$ using DESI+PantheonPlus data, respectively. All three models consistently show a deviation from CDDR at approximately $2\sigma$ confidence level, especially for the $P_3$ model, which has reached $2.4\sigma$. For further discussion on the $\sim 2\sigma$ deviation from the CDDR, please refer to Appendix~\ref{appendixB}. In contrast, if we adopt an $M_{\rm B}^{\rm B23}$ prior with $M_{\rm B}=-19.396\pm0.016$ mag, the constraint values of $\eta_0$ are $-0.035\pm0.056$ ($P_1$ model), $-0.070\pm0.124$ ($P_2$ model), and $-0.043\pm0.088$ ($P_3$ model), suggesting that no deviation from the CDDR is found in all models. This indicates that if a fixed $M_{\rm B}$ prior is adopted, a bias of approximately 0.2 mag of $M_{\rm B}$ can lead to a deviation of $2\sigma$ in CDDR, emphasizing the necessity of analysing the impact of the $M_{\rm B}$ prior on CDDR deviation. Furthermore, we can obtain that the absolute magnitude of the SN needs to be calibrated to approximately $-19.4$ mag in order for the CDDR to remain undeviated using the DESI and PantheonPlus data. 

Next, we present the 1D marginalized posterior distributions of $\eta_0$ for the $P_1$, $P_2$, and $P_3$ models, obtained using the DESI+DESY5 and SDSS+DESY5 data, as shown in Fig.~\ref{fig4}. Using the SDSS+DESY5 data, the constraint values of $\eta_0$ are $-0.035\pm0.074$, $-0.080\pm0.120$, and $-0.059\pm0.093$ for $P_1$, $P_2$, and $P_3$ models, respectively. Similarly, the constraint values of $\eta_0$ are $-0.052\pm0.066$ ($P_1$ model), $-0.090\pm0.110$ ($P_2$ model), and $-0.068\pm0.085$ ($P_3$ model) using DESI+DESY5 data, respectively. We find that these results consistently show no deviation from the CDDR in all parameterized models. Therefore, except for the case of adopting the $M_{\rm B}^{\rm D20}$ prior, we did not detect any deviation from CDDR in other cases, indicating that the current data are well consistent with CDDR.

Finally, it is necessary to compare the constraints on $\eta_0$ given by different observations to explore the capabilities of BAO and SN data. The constraints of $\eta_0$ using DESI+DESY5 are approximately 36\%, 56\%, and 47\% higher than those obtained using the SDSS eBOSS DR16 BAO dataset combined with the Pantheon SN sample, for $P_1$, $P_2$, and $P_3$ models, respectively \cite{Xu:2022zlm}. Similarly, the constraints are approximately 49\%, 71\%, and 72\% more restrictive than those obtained from six well-studied strong gravitational lensing systems and PantheonPlus SN data for three models \cite{Qi:2024acx}. Furthermore, these results are comparable to those inferred from compact radio quasars measurements and Pantheon SN samples \cite{Yang:2024icv}. Therefore, the latest BAO and SN data provide a more effective and accurate way to test CDDR.

\section{Conclusion}\label{sec4}

The CDDR is a fundamental relationship of modern cosmology, and its validation largely relies on precise distance measurements. In this work, we explore the potential deviation of the CDDR using three phenomenological parameterizations of the function $\eta(z)$, by combining BAO data from SDSS and DESI DR2 with SN data from PantheonPlus and DESY5. We also evaluate the impacts of the SN absolute magnitude on testing the CDDR.

When $M_{\rm B}$ is treated as a free parameter, we find a strong negative correlation between $\eta_0$ and $M_{\rm B}$, indicating that the uncertainty in $M_{\rm B}$ directly affects the inferred CDDR deviation. Meanwhile, all three parameterizations show no evidence for a deviation from CDDR using the SDSS, DESI, and PantheonPlus data. The best-fit values of $M_{\rm B}$ are slightly lower than the commonly adopted value of $-19.3$ mag, suggesting that the inclusion of a deviation parameter in CDDR can lead to a small shift in $M_{\rm B}$ from its conventional calibration. When fixing the $M_{\rm B}^{\rm D20}$ prior ($-19.230\pm0.040$ mag), we find a deviation from CDDR at the approximately $2\sigma$ level, especially in the $P_3$ model, where the deviation reaches the $2.4\sigma$ level. This deviation is due to a calibration tension between $M_{\rm B}^{\mathrm{D20}}$ and the Planck-calibrated $r_\mathrm{d}$, a mismatch that is closely connected to the $H_0$ tension. In contrast, the $M_{\rm B}^{\rm B23}$ prior ($-19.396\pm0.016$ mag) shows no deviation from CDDR in all models. This implies that a shift of approximately 0.2 mag in $M_{\rm B}$ can lead to a $2\sigma$ level deviation from CDDR. Finally, the joint analysis of DESI, SDSS, and DESY5 data maintains CDDR across all models.

In conclusion, our analysis has conducted a meticulous examination of the CDDR, revealing that the absolute magnitude $M_{\rm B}$ of SN has a significant impact on any deviation from the CDDR. Therefore, it is crucial in future to employ more data to calibrate $M_{\rm B}$ with higher accuracy and precision, thereby determining its value and re‐evaluating the CDDR.

\begin{figure*}[!htp]
\includegraphics[scale=0.35]{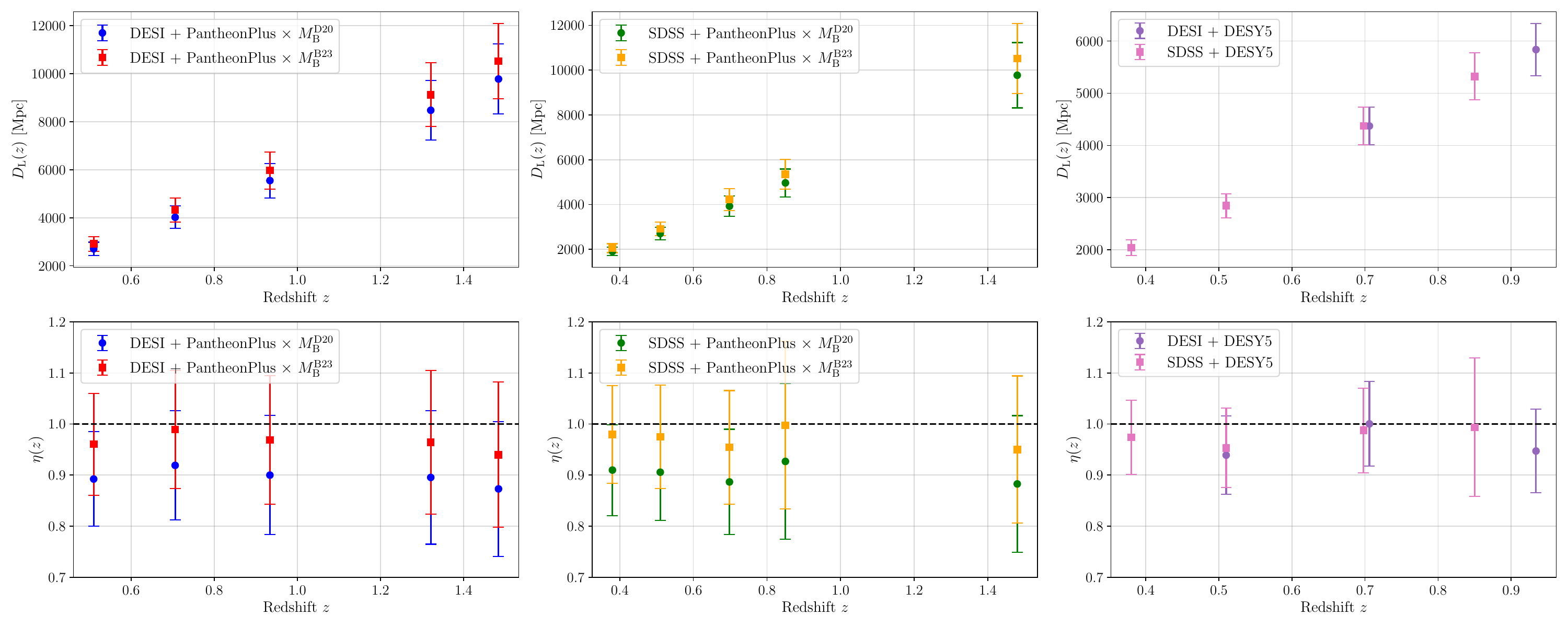}
\centering
\caption{\label{fig5}The Hubble diagram $D_\mathrm{L}(z)$ and the distance duality ratio $\eta(z)$ from SN and BAO data. Error bars represent the 1$\sigma$ measurement uncertainties. The ratio $\eta(z)$ is expected to be unity, indicated by the dashed line, if the CDDR holds. Notably, because DESY5 covers only a limited redshift range, the resulting $\eta(z)$ data comprise only three points for DESI and four points for SDSS.}
\end{figure*}

\section*{Acknowledgments}
We thank Tian-Yang Sun, Sheng-Han Zhou, Jia-Le Ling, Yi-Min Zhang, and Si-Ren Xiao for their helpful discussions. This work was supported by the National SKA Program of China (Grants Nos. 2022SKA0110200 and 2022SKA0110203), the National Natural Science Foundation of China (Grants Nos. 12533001, 12575049, 12473001, and 12205039), the China Manned Space Program (Grant No. CMS-CSST-2025-A02), and the National 111 Project (Grant No. B16009).

\textbf{Data Availability Statement} This manuscript has no associated data. [Author’s comment: All data analysed in this work is publicly available and references have been provided in Sect.~\ref{sec2}.]

\textbf{Code Availability Statement} This manuscript has no associated code/software. [Author’s comment: In this work, we utilized the publicly available software packages Cobaya and GetDist, with relevant references provided in Sect.~\ref{sec2}]

\appendix
% \section{\textcolor{blue}{The SN Hubble diagram $D_\mathrm{L}(z)$ and the distance duality ratio $\eta(z)$}}\label{appendixA}
\section{The SN Hubble diagram $D_\mathrm{L}(z)$ and the distance duality ratio $\eta(z)$}\label{appendixA}

In this appendix, we present the results of the luminosity distance $D_\mathrm{L}(z)$ reconstructed from PantheonPlus (considering two different $M_{\rm B}$ priors) and DESY5 SN data, and the distance duality ratio $\eta(z)$, constructed using $D_{\rm A}$ from BAO and $D_{\rm L}$ from SN, as shown in Fig.~\ref{fig5}. The blue data points corresponding to the combination DESI+PantheonPlus $\times$ $M_{\rm B}^{\rm D20}$ and the green data points for SDSS+PantheonPlus $\times$ $M_{\rm B}^{\rm D20}$ exhibit a noticeable deviation from the expected value $\eta(z) = 1$, suggesting a potential violation of the CDDR. In contrast, the red data points corresponding to the combination DESI+PantheonPlus $\times$ $M_{\rm B}^{\rm B23}$ and the orange data points for SDSS+PantheonPlus $\times$ $M_{\rm B}^{\rm B23}$ are in better agreement with $\eta(z) = 1$. These results underscore the limitations of testing the CDDR using calibration-dependent distance probes, particularly the dependence on the parameter $M_{\rm B}$. Additionally, the purple data points for DESI+DESY5 and the pink data points for SDSS+DESY5 are in close agreement, both being consistent with $\eta(z) = 1$.

\section{A brief discussion on the $\sim 2\sigma$ deviation from the CDDR}\label{appendixB}

To assess whether the $\sim2\sigma$ departure from the CDDR when adopting the $M_{\rm B}^{\mathrm{D20}}$ prior is driven by the assumed BAO sound horizon, we repeated the analysis fixing $r_{\rm d} = 138.3\ \mathrm{Mpc}$~\cite{Wang:2024rxm}. The constraint results of $\eta_0$ are reported in Table~\ref{tableA1}. The updated fits show that, for all parametrisations and for both SDSS+PantheonPlus and DESI+PantheonPlus data, the values of $\eta_0$ are consistent with zero at less than the $1\sigma$ level; accordingly, the previously reported $\sim2\sigma$ deviation disappears when a smaller $r_\mathrm{d}$ is adopted. This behaviour points to a calibration inconsistency between $M_{\rm B}^{\mathrm{D20}}$ and the Planck-calibrated $r_\mathrm{d}$, a mismatch that is also related to the $H_0$ tension.

\begin{table}[!htbp]
\centering
\caption{\label{tableA1} Constraint results of $\eta_0$ in the $P_1$, $P_2$, and $P_3$ models when $r_\mathrm{d}=138.3\ \mathrm{Mpc}$ is adopted~\cite{Wang:2024rxm}.}
\renewcommand{\arraystretch}{1.6}
\setlength{\tabcolsep}{4pt}
\resizebox{\columnwidth}{!}{%
\begin{tabular}{@{\extracolsep{\fill}}>{\rule{0pt}{2.5ex}}lcc}
\hline\hline
\textbf{Model}\rule{0pt}{3ex}
& SDSS + PantheonPlus $\times M_{\rm B}^{\rm D20}$
& DESI + PantheonPlus $\times M_{\rm B}^{\rm D20}$ \\
& $\eta_0$ & $\eta_0$ \\
\hline
$P_1$ model & $-0.043\pm 0.068$ & $-0.037\pm 0.052$ \\
$P_2$ model & $-0.080\pm0.125$ & $-0.080\pm 0.110$ \\
$P_3$ model & $-0.065\pm 0.089$ & $-0.068\pm 0.083$ \\
\hline\hline
\end{tabular}%
}
\end{table}

\bibliography{main}

\end{document}